\documentclass[12pt,cite,epsf,epsfig]{article}
\usepackage{epsfig}

\begin{document}

\author{S. Dev\thanks{dev5703@yahoo.com} and
Sanjeev Kumar\thanks{sanjeev3kumar@gmail.com}}
\title{\textbf{Two-Zero Symmetric Neutrino Mass Matrices\\ in Minimal Supersymmetric SO(10)}}
\date{Department of Physics, Himachal Pradesh University,
Shimla 171005, INDIA} \maketitle

\begin{abstract}
The phenomenological neutrino mass matrix for two-zero symmetric
texture has been obtained and used to rule out all possible
two-zero symmetric neutrino mass matrices obtained from Yukawa
couplings to $\mathbf{10}$ and $\mathbf{126}$ Higgs
representations within the framework of minimal supersymmetric
SO(10).
\end{abstract}

\section{Introduction}

The origin of fermion masses and mixings alongwith the related
problem of CP violation constitute a formidable challenge for
elementary particle physics. Leaving apart extremely small
neutrino masses, even the charged fermion mass hierarchy ranges
over at least five orders of magnitudes. Since the fermion masses
and the mixing angles are derived from the Yukawa couplings, which
are free parameters within the Standard Model (SM), these Yukawa
couplings must span several orders of magnitude to accommodate the
strongly hierarchical pattern of fermion masses and mixings.
However, the currently available data on fermion masses and mixing
are insufficient for an unambiguous reconstruction of fermion mass
matrices. To make matters worse, radiative corrections can obscure
the underlying structure. Thus, the existing data cannot, without
some additional assumptions, determine all the elements of the
Yukawa coupling matrices for quarks and leptons. Some of these
assumptions, invoked to restrict the form of fermion mass matrices
include the presence of texture zeros \cite{FGM paper},
requirement of zero determinant \cite{zero determinant} and zero
trace condition \cite{zero trace} to name just a few. The main
motivation for invoking different mass matrix ansatze is to relate
fermion masses and mixing angles in a testable manner which
reduces the number of free parameters in the Yukawa sector of SM.
The recent evidence for non-zero neutrino masses and mixings leads
to a further proliferation of free parameters in the Yukawa
sector. In the absence of a significant breakthrough in the
theoretical understanding of fermion flavors, the phenomenological
approaches are bound to play a crucial role in interpreting new
experimental data on quark and lepton mixing. These approaches are
expected to provide useful hints toward unaraveling the dynamics
of fermion mass generation, CP violation and identification of
possible underlying symmetries of the fermion flavors from which
viable models of fermion mass generation and flavor mixing could,
hopefully, be constructed.

The strong fermion mass hierarchy should be apparent in the
fermion mass matrices themselves with the contribution of smaller
elements to physical masses and mixing angles expected to be
negligibly small. Thus, these elements can, effectively, be
neglected and replaced by zeros: the so-called texture zeros.
However, the current neutrino oscillation data are consistent only
with a limited number of texture schemes \cite{FGM paper}.
Specifically, the available neutrino oscillation data disallow all
neutrino mass matrices with three or more texture zeros \cite{FGM
paper} in the flavor basis. The texture zeros in different
positions in the neutrino mass matrix, in particular, and fermion
mass matrices, in general, could be the consequence of some
underlying symmetry. Such universal textures of fermion mass
matrices can be realized within the framework of Grand Unified
Theories (GUTs). Though Grand Unification on its own does not shed
any light on the flavor problem, the GUTs provide the optimal
framework in which possible solutions to the flavor problem could
be embedded. As mentioned earlier, texture zeros in different
positions of the fermion mass matrices could result because of an
underlying symmetry. Grand Unified  models attempt to explain the
masses and mixings in both the quark and lepton sectors
simultaneously. The textures for the mass matrices obtained in
these models can either be assumed at the very outset or can be
derived from the observed mixing matrix in the flavor basis.
Alternatively, the textures of the mass matrix can be obtained by
embedding some family symmetry within the chosen Grand Unification
group. One particularly interesting class of models is that based
upon the SO(10) grand unification group. There are two kinds of
minimal models in this class: those based upon Higgs
representation with dimension $\mathbf{10}$, $\mathbf{126}$,
$\mathbf{\overline{126}}$ and possibly also $\mathbf{120}$ and/or
$\mathbf{210}$ and those based upon $\mathbf{10}$, $\mathbf{16}$,
$\mathbf{\overline{16}}$, and $\mathbf{45}$ representations. The
former choice, generally, has symmetric and/or antisymmetric
texture mass matrices while the latter type generally imply
lopsided mass matrices for the down type quarks and charged
leptons. In the present work, we restrict ourselves to the former
choice \textit{i.e.} we adopt the so called symmetric two-zero
texture for the up-type quark mass matrices within the SUSY SO(10)
GUT framework. Within this framework, we not only have a relation
between down type quark mass matrices ($M_d$) and charged lepton
mass matrices ($M_l$) but also a relation between the up-type
quark mass matrices ($M_u$) and Dirac neutrino mass matrices
($M_{\nu_{D}}$). Thus, once we fix the representation of the Higgs
field corresponding to each element, $M_l$ and $M_{\nu_{D}} $ are
uniquely determined from $M_d$ and $M_u$ respectively.

Neutrino mass matrices with two texture zeros in the charged
lepton basis have only one degree of freedom \cite{a1a2 paper, 2t
paper}. The presence of two texture zeros in the neutrino mass
matrix imply four conditions on the nine free parameters of the
model. The four parameters out of the remaining five parameters
are determined by the neutrino data for the values of two
squared-mass differences $\Delta m^2_{12}$ and $\Delta m^2_{23}$
and two mixing angles $\theta_{12}$ and $\theta_{23}$. So, we are
left with only one free parameter in the neutrino mass matrix
\cite{a1a2 paper, 2t paper}. However, we have one more
experimental measurement by the CHOOZ experiment which establishes
an upper bound on the third mixing angle $\theta_{13}$. A lower
bound on $\theta_{13}$ is inherent in the nature of two texture
zero neutrino mass matrices \cite{a1a2 paper,cpv paper}. So, even
the remaining one degree of freedom is constrained. In fact, the
neutrino mass matrix with two texture zeros can be completely
determined \cite{Desai} with the help of the presently available
neutrino data for the seven allowed texture zero schemes of
Frampton, Glashow and Marfatia \cite{FGM paper}. Therefore, the
present neutrino data, which is sufficient to determine the
neutrino mass matrix with two texture zeros, will be able to
overrule various neutrino mass models with two texture zeros if
the model parameters are known and to constrain the model
parameters if they are unknown.

In the class of GUT models under consideration, the mass matrices
$M_{\nu_ D}$ for the Dirac neutrinos which gives rise to neutrino
masses via see-saw mechanism can be taken identical to $M_{u}$
except for the accompanying CG coefficients which depend upon the
representations of the coupling Higgs field \cite{Bando and Obara
1, Bando and Obara 2}. Therefore, these mass models predict the
neutrino mass matrix at GUT scale. This theoretical mass matrix
should be consistent with the phenomenological mass matrix at weak
scale calculated for the texture scheme consistent with the model.
However, the neutrino mass matrix calculated at GUT scale has to
be run down to the weak scale before comparing it with the
neutrino mass matrix calculated phenomenologically from its
texture scheme from the neutrino data at the weak scale. The
effect of renormalization group (RG) running will be small for the
neutrino mass matrices with normal hierarchy. However, it can be
significantly large for the mass matrices with inverted or
quasi-degenerate hierarchy.

In the present work, we first derive the phenomenological neutrino
mass matrix for a particular texture scheme proposed by Frampton,
Glashow and Marfatia \cite{FGM paper} and confront it with a class
of two-zero symmetric texture GUT models based upon minimal
supersymmetric SO(10) \cite{Bando and Obara 1, Bando and Obara 2}.

\section{Phenomenological neutrino mass matrix with a two zero
symmetric texture}

We take the neutrino mass matrix to be the real symmetric matrix
with two texture zeros:
\begin{equation}
M_{\nu}=\left( \begin{array}{ccc} 0 & A & 0\\ A & D & B \\0 & B &
C\\
\end{array} \right)
\end{equation}
where
\begin{equation}
 \begin{array}{c}
C=m_1-m_2+m_3-D, ~~~ A^2=m_1m_2m_3/C,\\
B^2=(m_3+m_1-D)(m_3-m_2-D)(m_2-m_1+D)/C \end{array}
\end{equation}
since we are interested in obtaining only the relative magnitudes
of the neutrino mass matrix elements without phases or signs. This
is the texture $\mathcal{A}_2$ of Frampton, Glashow and Marfatia
\cite{FGM paper}. The eigenvalues of $M_{\nu}$ are $m_1$, $-m_2$
and $m_3$. The real orthogonal matrix $O$ which diagonalizes the
neutrino mass matrix $M$ according to the relation
\begin{equation}
O^T M O=diag(m_1,-m_2,m_3)
\end{equation}
is given by
\begin{equation}
 \left.\begin{array}{c}

O_{11}=\sqrt{\frac{m_2m_3(m_3-m_2-D)}{(m_1-m_2+m_3-D)
(m_1+m_2)(m_3-m_1)}}
\\

O_{12}=\sqrt{\frac{m_3m_1(m_3+m_1-D)}{(m_1-m_2+m_3-D)
(m_1+m_2)(m_3+m_2)}}   \\

O_{13}=\sqrt{\frac{m_1m_2(m_2-m_1+D)}{(m_1-m_2+m_3-D)
(m_2+m_3)(m_3-m_1)}}   \\

O_{21}=\sqrt{\frac{m_1(m_3-m_2-D)}{(m_1+m_2)(m_3-m_1)}}   \\

O_{22}=\sqrt{\frac{m_2(m_3+m_1-D)}{(m_2+m_3)(m_1+m_2)}} \\

O_{23}=\sqrt{\frac{m_3(m_2-m_1+D)}{(m_3-m_1)(m_2+m_3)}} \\

O_{31}=\sqrt{\frac{m_1(m_2-m_1+D)(m_3+m_1-D)}{(m_1-m_2+m_3-D)
(m_2+m_1)(m_3-m_1)}}   \\

O_{32}=\sqrt{\frac{m_2(m_2-m_1+D)(m_3-m_2-D)}{(m_1-m_2+m_3-D)
(m_3+m_2)(m_1+m_2)}}   \\

O_{33}=\sqrt{\frac{m_3(m_3-m_2-D)(m_3+m_1-D)}{(m_1-m_2+m_3-D)
(m_2+m_3)(m_3-m_1)}}

\end{array}  \right\}
\end{equation}
where we have given only the magnitudes of the elements of $O$.
This diagonalization scheme is taken from \cite{Ahuja} where the
references to earlier works can, also, be found.

Many interesting relations can be extracted from the above
expressions for the elements of the neutrino mass matrix and its
eigenvalues. The two mass ratios $\frac{m_1}{m_2}$ and
$\frac{m_1}{m_3}$ can be written as
\begin{equation}
\frac{m_1}{m_2}=\frac{O_{12}O_{21}}{O_{11}O_{22}}
\end{equation}
and
\begin{equation}
\frac{m_1}{m_3}=\frac{O_{13}O_{21}}{O_{11}O_{23}}.
\end{equation}
The elements $A$, $B$ and $C$ are given by
\begin{equation}
A=\sqrt{m_1m_2\frac{O_{11}O_{12}}{O_{21}O_{22}}},
\end{equation}
\begin{equation}
B=\frac{(m_1+m_2)(m_2+m_3)(m_3-m_1)}{m_1m_3} O_{12}O_{21}O_{23}
\end{equation}
and
\begin{equation}
C=m_3\frac{O_{21}O_{22}}{O_{11}O_{12}}.
\end{equation}
Therefore, the ratios $\frac{A}{C}$ and $\frac{B}{C}$ become
\begin{equation}
\frac{A}{C}=\frac{O_{11}O_{12}O_{13}}{O_{21}O_{22}O_{23}}
\end{equation}
and
\begin{equation}
\frac{B}{C}=O_{11}O_{12}O_{13}\frac
{(m_1+m_2)(m_2+m_3)(m_3-m_1)}{m_1m_2m_3}.
\end{equation}
Another important relation is
\begin{equation}
O_{13}=\frac{\sqrt{m_1m_2}}{m_3}
O_{23}\sqrt{\frac{O_{11}O_{12}}{O_{21}O_{22}}}.
\end{equation}
Eqs. (5)-(12) are exact and have been derived from Eqs. (4). These
relations imply very interesting consequences when written
approximately as Taylor series in the powers of $s_{13}$ coming
from the orthogonal mixing matrix $O$ which can be written as
\begin{equation}
O=\left(
\begin{array}{ccc}
c_{12}c_{13} & s_{12}c_{13} & s_{13} \\
-s_{12}c_{23}-c_{12}s_{23}s_{13} & c_{12}c_{23}-s_{12}s_{23}s_{13}
& s_{23}c_{13} \\ s_{12}s_{23}-c_{12}c_{23}s_{13} &
-c_{12}s_{23}-s_{12}c_{23}s_{13} & c_{23}c_{13}
\end{array}
\right)
\end{equation}
where the symbols have their usual meaning. We substitute these
elements in equations (5)-(12) and expand the results in the
powers of $s_{13}$. The mass ratios in Eqs. (5) and (6) are given
by
\begin{equation}
\frac{m_1}{m_2}=\frac{s^2_{12}}{c^2_{12}}+\mathcal{O}(s_{13})
\end{equation}
and
\begin{equation}
\frac{m_1}{m_3}=\frac{s_{12}c_{23}s_{13}}{c_{12}s_{23}}
+\mathcal{O}(s^2_{13})
\end{equation}
which are consistent with our earlier results \cite{a1a2 paper, 2t
paper}. The mass ratio $\frac{m_1}{m_2}$ is much smaller than one.
Therefore, the neutrino mass matrix considered here has a normal
hierarchy. This fact is important since the neutrino oscillation
parameters change very little in the RG evolution from the weak
scale to GUT scale for the normal hierarchy. The matrix elements
$A$, $B$ and $C$ are given by
\begin{equation}
A=\frac{\sqrt{m_1m_2}}{c_{23}}+\mathcal{O}(s_{13}),
\end{equation}
\begin{equation}
B=m_3c_{23}s_{23}+\mathcal{O}(s_{13})
\end{equation}
and
\begin{equation}
C=m_3c_{23}^2+\mathcal{O}(s_{13}).
\end{equation}
Moreover, the element $D$ can be expressed as
\begin{equation}
D=m_3s_{23}^2+\mathcal{O}(s_{13}).
\end{equation}
The ratios $\frac{A}{C}$ and $\frac{B}{C}$ can be written as
\begin{equation}
\frac{A}{C}=\frac{s_{13}}{c_{23}^2s_{23}}+\mathcal{O}(s^2_{13})
\end{equation}
and
\begin{equation}
\frac{B}{C}=1+\mathcal{O}(s_{13}).
\end{equation}
Eq. (12) for $O_{13}$ gives
\begin{equation}
s_{13}=\frac{\sqrt{m_1m_2}}{m_3}+\mathcal{O}(s^2_{13}).
\end{equation}
This equation is consistent with the analytical results obtained
by Desai \textit{et al.} \cite{Desai}. Eq. (14) can be used to
obtain the relation
\begin{equation}
m_1= s^2_{12}\sqrt{\frac{\Delta m^2_{12}}{\cos 2 \theta_{12}}}
\end{equation}
as shown in our earlier work \cite{mee paper}. Using Eq. (23) and
neglecting the higher order terms in $s_{13}$, Eq. (22) can be
written as
\begin{equation}
s_{13}\approx\left(\frac{\Delta m^2_{12}}{\Delta
m^2_{13}}\right)^{\frac{1}{2}}\frac{s_{12}c_{12}}
{\sqrt{c^2_{12}-s^2_{12}}}.
\end{equation}
For the tri-bi-maximal value \cite{TBM} $s_{12}^2=\frac{1}{3}$,
the trigonometric factor in Eq. (24) is given by
\begin{equation}
\frac{s_{12}c_{12}} {\sqrt{c^2_{12}-s^2_{12}}} =
\sqrt{\frac{2}{3}}.
\end{equation}
Substituting the present best fit values of the oscillation
parameters in Eq. (24) ($\Delta m^2_{12}=7.9\times10^{-5}$,
$\Delta m^2_{13}=2.4\times10^{-3}$, and $s_{12}=0.3$
\cite{fogli}), we find that $\theta_{13}= 7.3^o$ which is
consistent with the earlier estimates \cite{a1a2 paper,2t paper}
for this mass matrix texture.

Using the above leading order estimates of the elements $A$, $B$,
$C$ and $D$, the neutrino mass matrix given in Eq. (1) can be
written as
\begin{equation}
M=\frac{m_3}{2}\left( \begin{array}{ccc} 0 & \epsilon & 0\\
\epsilon &1+\mathcal{O}(\epsilon) & 1+\mathcal{O}(\epsilon)
\\0 & 1+\mathcal{O}(\epsilon) & 1+\mathcal{O}(\epsilon)\\
\end{array} \right)
\end{equation}
where the parameter $\epsilon$ is given by
\begin{equation}
\epsilon=2\sqrt{2}s_{13}
\end{equation}
for maximal atmospheric mixing. For the present best fit values of
the neutrino oscillation parameters, $\epsilon\sim 0.36$. So, the
hierarchical structure of neutrino mass matrix should be as
follows:
\begin{equation}
M\sim\frac{1}{2}\sqrt{\Delta m^2_{13}}\left( \begin{array}{ccc} 0
&0.36 & 0\\0.36 &1 & 1\\0 & 1 & 1\\
\end{array} \right)
\end{equation}

\section{A minimal supersymmetric SO(10) realization of the
two-zero symmetric texture}

The phenomenological neutrino mass matrix obtained here has to be
confronted with the neutrino mass matrices given by various
neutrino mass models. As an example, we confront a class of
minimal supersymmetric SO(10) GUT models with two-zero symmetric
texture \cite{Bando and Obara 1, Bando and Obara 2} with the
phenomenological mass matrix obtained above. In this class of
models, each element of $M_u$ is dominated by contribution either
from $\mathbf{10}$ or $\mathbf{126}$ Higgs representation and the
Yukawa couplings of Dirac neutrinos $\nu_{D}$ are that of
corresponding up quarks multiplied by a factor of 1 or -3
respectively. Therefore, the Dirac neutrino mass matrix for this
class of models can be written as
\begin{equation}
M_{\nu_ D}= \left( \begin{array}{ccc} 0&a&0\\ a&b&c\\0&c&d \\
\end{array} \right)m_t
\end{equation}
which is related to the up-quark mass matrix $M_u$
\begin{equation}
M_u=\left( \begin{array}{ccc} 0&a_u&0\\ a_u&b_u&c_u\\0&c_u&1 \\
\end{array} \right)m_t
\end{equation}
through the relations
\begin{equation}
 a=x_{12}a_u, ~
 b=x_{22}b_u, ~
 c=x_{23}c_u, ~
 d=x_{33}
\end{equation}
where
\begin{equation}
a_u=\frac{\sqrt{m_u m_c}}{m_t}, ~ b_u=\frac{m_c}{m_t}, ~
c_u=\sqrt{\frac{m_u}{m_t}}
\end{equation}
and $m_u$, $m_c$ and $m_t$ are the masses of the up quarks u, c
and t at the GUT scale. This mass matrix is consistent with the
quark masses and mixing data. The coefficients $x_{12}$, $x_{22}$,
$x_{23}$ and $x_{33}$ are the CG factors which connect the
respective elements of $M_u$ and $M_{\nu_ D}$ and are equal to 1
or -3 depending upon whether the corresponding Higgs
representation is \textbf{10} or \textbf{126}. The Higgs
representation for $M_u$ which has the best agreement with the
data is given by \cite{Bando and Obara 1}
\begin{equation}
M_u~~~:~~~\left( \begin{array}{ccc} 0&\mathbf{126}&0\\
\mathbf{126}&\mathbf{10}&\mathbf{10}\\0&\mathbf{10}&\mathbf{126}
\\
\end{array} \right).
\end{equation}
For this $M_u$, the CG coefficients are given by
\begin{equation}
x_{12}=-3, ~ x_{22}=1, ~
 x_{23}=1, ~
 x_{33}=-3.
\end{equation}
The link with the neutrino mass matrix is established through the
see-saw mechanism using the right handed neutrino mass matrix
matrix $M_R$:
\begin{equation}
M_R=\left( \begin{array}{ccc} 0&r m_R&0\\r m_R&0&0\\0&0&m_R \\
\end{array} \right).
\end{equation}
This particular form of $M_R$ is consistent with the \textbf{126}
dimensional representation for Higgs field which couples with up
quarks. It gives two degenerate right handed neutrino states of
mass $rm_R$ and a third neutrino state of mass $m_R$. Thus, $r$ is
the ratio of two right handed neutrino mass scales. The neutrino
mass matrix $M_{\nu}$ can be obtained from the right-handed
Majorana mass matrix ($M_R$) and the Dirac mass matrix ($M_{\nu_
D}$) via the see-saw relation
\begin{equation}
M_{\nu}=M^T_{\nu_ D} M^{-1}_R M_{\nu_ D}
\end{equation}
and is given by
\begin{equation}
M_{\nu}= \left( \begin{array}{ccc} 0&\frac{a^2}{r}&0\\
\frac{a^2}{r}&\frac{2ab}{r}+c^2&c(\frac{a}{r}+1)\\0&c(\frac{a}{r}+1)
&d^2\\
\end{array} \right)\frac{m_t^2}{m_R}.
\end{equation}
The above neutrino mass matrix, to a very good approximation, can
be written as \cite{Bando and Obara 1,Bando and Obara 2}
\begin{equation}
M_{\nu}=\left(\begin{array}{ccc}
 0   & \frac{a^2}{r} & 0\\
 \frac{a^2}{r} &  \frac{2ab}{r} &  \frac{ac}{r}\\
 0 &  \frac{ac}{r} & d^2\\
\end{array}\right) \frac{m_t^2}{m_R}.
\end{equation}

It will be convenient to parameterize $M_{\nu}$ in the following
way \cite{Bando and Obara 1,Bando and Obara 2}
\begin{equation}
M_{\nu}=\left(\begin{array}{ccc}
 0   &\beta & 0\\
 \beta &  \alpha &  h\\
 0 &  h & 1\\
\end{array}\right)  \frac{d^2 m_t^2}{m_R}
\end{equation}
where
\begin{equation}
\alpha=\frac{2 a b}{r d^2}=\frac{x_{12}x_{22}}{x_{33}^2}
\left(\frac{2 a_u b_u}{r}\right),
\end{equation}
\begin{equation}
\beta=\frac{a^2}{r d^2}=\frac{x_{12}^2}{x_{33}^2}
\left(\frac{a_u^2}{r}\right)
\end{equation}
and
\begin{equation}
h=\frac{a c}{r d^2}=\frac{x_{12}x_{23}}{x_{33}^2} \left(\frac{a_u
c_u}{r}\right)  .
\end{equation}
Since, $\alpha$, $\beta$ and $h$ are all proportional to
$\frac{1}{r}$, we can eliminate $r$ and work in term of the ratios
$\frac{\beta}{\alpha}$ and $\frac{\alpha}{h}$ which are given by
\begin{equation}
\frac{\beta}{\alpha}=\frac{x_{12}}{x_{23}}\left(\frac{1}{2}
\sqrt{\frac{m_u}{m_c}} \right)
\end{equation}
and
\begin{equation}
\frac{\alpha}{h}=\frac{x_{22}}{x_{23}} \left(
2\frac{m_c}{\sqrt{m_u m_t}} \right) .
\end{equation}

The ratio $\frac{\alpha}{h}$ is found to be of the order of unity
and is in agreement with the value obtained from the
phenomenological neutrino mass matrix [Eq. (26) and Eq. (28)].
However, this ratio depends on the top quark mass $m_t$ which has
rather large uncertainty. Moreover, the quark masses have to be
run to the GUT scale and the RG running effects are significantly
large for this ratio. Therefore, this ratio cannot serve as a good
criterion to test this model. On the other hand, the ratio
$\frac{\beta}{\alpha}$ depends on the relatively better known
quark masses $m_u$ and $m_c$ and the RG running effects in
calculating this ratio at the GUT scale are quite small. In fact,
this ratio is quite stable against the various input SUSY
parameters (like $\tan^2\beta$ and threshold corrections) which
affect the RG evolution  \cite{Ross and Serna}. Substituting the
value of the quark mass ratio $\frac{m_u}{m_c}$ at GUT scale
\cite{Ross and Serna}:
\begin{equation}
\frac{m_u}{m_c}=0.0027\pm0.0006,
\end{equation}
we find that the model prediction for the ratio
$\frac{\beta}{\alpha}$ is about $0.078\pm0.009$.

The ratio $\frac{\beta}{\alpha}$ can, also, be obtained from the
phenomenological neutrino mass matrix since
\begin{equation}
\frac{\beta}{\alpha}= \frac{A}{D}=
\frac{A}{C}+\mathcal{O}(s_{13})\sim\epsilon
\end{equation}
and hence $\frac{\beta}{\alpha}\sim 0.4$. This value has to be
evolved to the GUT scale by the RG running. However, it can be
seen that this ratio is quite stable against RG evolution which
can lower this ratio by a factor of about $\frac{9}{10}$ at the
most when going from the weak scale to the GUT scale. This can be
seen from the relations \cite{Bando and Obara 3}
\begin{equation}
\alpha\rightarrow \frac{\alpha}{(1-\epsilon_{\mu})^2},~\beta
\rightarrow \frac{\beta}{(1-\epsilon_e)(1-\epsilon_{\mu})},~
h\rightarrow \frac{h}{(1-\epsilon_{\mu})}
\end{equation}
where the sign `$\rightarrow$' denotes the RG evolution from the
weak scale to the GUT scale and $\epsilon_{e}$ and
$\epsilon_{\mu}$ are the RG parameters which can be at the most
$0.1$. Hence, the smallest possible value of the ratio
$\frac{\beta}{\alpha}$ at the GUT scale is approximately $0.3$
which is still larger than the model value for this ratio by a
factor of about $4$. This disagreement between the
phenomenological values of the ratio $\frac{\beta}{\alpha}$
calculated from the neutrino data and the values predicted by the
GUT model can not be reconciled even if the errors of the neutrino
oscillation data are taken into account. We have done the exact
numerical calculation and found that $\frac{\beta}{\alpha} =
0.36\pm0.05$. This value differs from the model value by more than
$5$ $\sigma$ C.L.. Therefore, the two-zero symmetric texture of
the neutrino mass matrix based upon the minimal supersymmetric
SO(10) GUT is ruled out by more than $5$ $\sigma$ C.L..

The neutrino mass matrix $M_{\nu}$ given in Eq. (39) corresponds
to the up-quark mass matrix $M_u$ with the Higgs representation
given in Eq. (33) which is only one of the possible Higgs
representations. $M_u$, a symmetric matrix with two texture zeros,
contains four independent elements which can couple with either
\textbf{10} or \textbf{126} dimensional Higgs fields. There are 16
such representations listed in Table 1 where we have adopted the
classification scheme of \cite{Bando and Obara 2} for comparison.
Only eight textures out of the total sixteen were found to be
allowed in earlier studies \cite{Bando and Obara 2}. Just like the
basic texture based upon the Higgs representation given in Eq.
(33) for $M_u$, the other fifteen possible textures with different
possible Higgs representation for $M_u$ are also ruled out by the
phenomenological values for the ratios $\frac{\beta}{\alpha}$ and
$\frac{\alpha}{h}$. The representation chosen for $M_u$ in Eq.
(33) is $\mathcal{S}_1$ and so we can call the $M_{\nu}$ obtained
in Eq. (39) to be of type $\mathcal{S}_1$. Similarly, one can
calculate the neutrino mass matrix for all the sixteen Higgs field
representations of $M_u$ which will differ only in the values of
the CG factors $x_{12}$, $x_{22}$, $x_{23}$ and $x_{33}$. These
factors for all the sixteen type of neutrino mass matrices have
been tabulated in Table 2 where we also calculate the parameters
$\alpha$, $\beta$ and $h$ in the units of $\left(\frac{2 a_u
b_u}{r}\right)$, $\left(\frac{a_u^2}{r}\right)$ and
$\left(\frac{a_u c_u}{r}\right)$ respectively and their ratios
$\frac{\beta}{\alpha}$ and $\frac{\alpha}{h}$ in the units of
$\left(\frac{1}{2} \sqrt{\frac{m_u}{m_c}} \right)$ and $\left(
2\frac{m_c}{\sqrt{m_u m_t}} \right)$ respectively. In these units,
the ratio $\frac{\alpha}{h}$ should be about $1$ and the ratio
$\frac{\beta}{\alpha}$ should be about $15$ to explain the present
neutrino data. It can be seen that the ratio $\frac{\alpha}{h}$ is
$1$ in agreement with its phenomenological value only for the
categories $\mathcal{S}$, $\mathcal{A}$ and $\mathcal{B}$. These
categories were favored by the experimental data in some earlier
studies \cite{Bando and Obara 1,Bando and Obara 2}. However, the
ratio $\frac{\beta}{\alpha}$ is approximately smaller by the
factors of $5$, $15$ and $45$ from its phenomenological value for
the classes $\mathcal{S}$, $\mathcal{A}$ and $\mathcal{B}$,
respectively, ruling out these categories. The categories
$\mathcal{C}$ and $\mathcal{F}$ are ruled out in a straightforward
manner since they do not reproduce either of the two ratios
correctly. Hence, the entire class of two-zero symmetric texture
of neutrino mass matrices is ruled out.

\section{Conclusions}

In conclusion, we have obtained the phenomenological neutrino mass
matrix for the two-zero symmetric texture and found that it is not
compatible with all possible two-zero symmetric texture neutrino
mass matrices obtained from the Yukawa couplings with
$\mathbf{10}$ and $\mathbf{126}$ Higgs representations within the
framework of minimal supersymmetric SO(10). Therefore, the
theoretical motivation for two-zero symmetric texture of the
neutrino mass matrix, no longer, exists.

\pagebreak

\begin{table}
\begin{center}
\begin{tabular}{|cc|cc|}
\hline

&&&\\

$\mathcal{S}_1~:~~$ & $\left( \begin{array}{ccc}
0&\mathbf{126}&0\\
\mathbf{126}&\mathbf{10}&\mathbf{10}\\0&\mathbf{10}
&\mathbf{126}\\\end{array} \right)$ &

$\mathcal{S}_2~:~~$ & $\left(\begin{array}{ccc} 0&\mathbf{126}&0\\
\mathbf{126}&\mathbf{10}&\mathbf{10}\\0&\mathbf{10}&\mathbf{10}
\\\end{array}
\right)$\\

&&&\\ \hline \hline &&&\\

$\mathcal{A}_1~:~~$ & $\left( \begin{array}{ccc}
0&\mathbf{126}&0\\
\mathbf{126}&\mathbf{126}&\mathbf{126}\\0&\mathbf{126}&\mathbf{126}
\\\end{array}
\right)$ &

$\mathcal{A}_2~:~~$ & $\left(\begin{array}{ccc} 0&\mathbf{126}&0\\
\mathbf{126}&\mathbf{126}&\mathbf{126}\\0&\mathbf{126}&\mathbf{10}\\\end{array}
\right)$ \\

$\mathcal{A}_3~:~~$ & $\left( \begin{array}{ccc} 0&\mathbf{10}&0\\
\mathbf{10}&\mathbf{10}&\mathbf{10}\\0&\mathbf{10}&\mathbf{126}\\\end{array}
\right)$ &

$\mathcal{A}_4~:~~$ & $\left(\begin{array}{ccc} 0&\mathbf{10}&0\\
\mathbf{10}&\mathbf{10}&\mathbf{10}\\0&\mathbf{10}&\mathbf{10}\\\end{array}
\right)$
\\

&&&\\ \hline \hline &&&\\

$\mathcal{B}_1~:~~$ & $\left( \begin{array}{ccc} 0&\mathbf{10}&0\\
\mathbf{10}&\mathbf{126}&\mathbf{126}\\0&\mathbf{126}&\mathbf{126}\\\end{array}
\right)$ &

$\mathcal{B}_2~:~~$ & $\left(\begin{array}{ccc} 0&\mathbf{10}&0\\
\mathbf{10}&\mathbf{126}&\mathbf{126}\\0&\mathbf{126}&\mathbf{10}\\\end{array}
\right)$ \\

&&&\\ \hline \hline &&&\\

$\mathcal{C}_1~:~~$ & $\left( \begin{array}{ccc}
0&\mathbf{126}&0\\
\mathbf{126}&\mathbf{10}&\mathbf{126}\\0&\mathbf{126}&\mathbf{126}\\\end{array}
\right)$ &

$\mathcal{C}_2~:~~$ & $\left(\begin{array}{ccc} 0&\mathbf{10}&0\\
\mathbf{10}&\mathbf{10}&\mathbf{126}\\0&\mathbf{126}&\mathbf{126}\\\end{array}
\right)$ \\

$\mathcal{C}_3~:~~$ & $\left( \begin{array}{ccc} 0&\mathbf{10}&0\\
\mathbf{10}&\mathbf{10}&\mathbf{126}\\0&\mathbf{126}&\mathbf{10}\\\end{array}
\right)$ &

$\mathcal{C}_4~:~~$ & $\left(\begin{array}{ccc} 0&\mathbf{126}&0\\
\mathbf{126}&\mathbf{10}&\mathbf{126}\\0&\mathbf{126}&\mathbf{10}\\\end{array}
\right)$ \\

&&&\\ \hline \hline &&&\\

$\mathcal{F}_1~:~~$ & $\left( \begin{array}{ccc}
0&\mathbf{126}&0\\
\mathbf{126}&\mathbf{126}&\mathbf{10}\\0&\mathbf{10}&\mathbf{126}\\\end{array}
\right)$ &

$\mathcal{F}_2~:~~$ & $\left(\begin{array}{ccc} 0&\mathbf{10}&0\\
\mathbf{10}&\mathbf{126}&\mathbf{10}\\0&\mathbf{10}&\mathbf{126}\\\end{array}
\right)$ \\

$\mathcal{F}_3~:~~$ & $\left( \begin{array}{ccc} 0&\mathbf{10}&0\\
\mathbf{10}&\mathbf{126}&\mathbf{10}\\0&\mathbf{10}&\mathbf{10}\\\end{array}
\right)$ &

$\mathcal{F}_4~:~~$ & $\left(\begin{array}{ccc} 0&\mathbf{126}&0\\
\mathbf{126}&\mathbf{126}&\mathbf{10}\\0&\mathbf{10}&\mathbf{10}\\\end{array}
\right)$ \\

 &&&\\ \hline
\end{tabular}
\caption{The sixteen possible Higgs representations for $M_u$ as
classified in Ref. \cite{Bando and Obara 2}.}
\end{center}
\end{table}

\begin{table}
\begin{center}
\begin{tabular}{|c|cccc|ccc|cc|}
\hline

&&&&&&&&&\\

 & $x_{12}$ & $x_{22}$ & $x_{23}$ & $x_{33}$ &
 $\alpha=\frac{x_{12}x_{22}}{x^2_{33}}$ &
 $\beta=\frac{x_{12}^2}{x^2_{33}}$ &
$h=\frac{x_{12}x_{23}}{x^2_{33}}$ &
$\frac{\beta}{\alpha}=\frac{x_{12}}{x_{23}}$ & $\frac{\alpha
}{h}=\frac{x_{22}}{x_{23}}$
\\

&&&&&&&&&\\ \hline \hline

 $\mathcal{S}_1$ & -3 & 1 & 1 & -3 & -1/3 & 1 & -1/3 & -3 & 1 \\
 $\mathcal{S}_2$ & -3 & 1 & 1 & 1 & -3 & 9 & -3 & -3 & 1 \\  \hline \hline
 $\mathcal{A}_1$ & -3 & -3 & -3 & -3 & 1 & 1 & 1 & 1 & 1 \\
 $\mathcal{A}_2$ & -3 & -3 & -3 &  1 & 9 & 9 & 9 & 1 & 1 \\
 $\mathcal{A}_3$ & 1 & 1 & 1 & -3 & 1/9 & 1/9 & 1/9 & 1 & 1 \\
 $\mathcal{A}_4$ & 1 & 1 & 1 & 1 & 1 & 1 & 1 & 1 & 1 \\     \hline \hline
 $\mathcal{B}_1$ & 1 & -3 & -3 & -3 & -1/3 & 1/9 & -1/3 & -1/3 & 1 \\
 $\mathcal{B}_2$ & 1 & -3 & -3 & 1 & -3 & 1 & -3 & -1/3 & 1 \\  \hline \hline
 $\mathcal{C}_1$ & -3 & 1 & -3 & -3 & -1/3 & 1 & 1 & -3 & -1/3 \\
 $\mathcal{C}_2$ & 1 & 1 & -3 & -3 & 1/9 & 1/9 & -1/3 & 1 & -1/3 \\
 $\mathcal{C}_3$ & 1 & 1 & -3 & 1 & 1 & 1 & -3 & 1 & -1/3 \\
 $\mathcal{C}_4$ & -3 & 1 & -3 & 1 & -3 & 9 & 9 & -3 & -1/3 \\  \hline \hline
 $\mathcal{F}_1$ & -3 & -3 & 1 & -3 & 1 & 1 & -1/3 & 1 & -3 \\
 $\mathcal{F}_2$ & 1 & -3 & 1 & -3 & -1/3 & 1/9 & 1/9 & -1/3 & -3 \\
 $\mathcal{F}_3$ & 1 & -3 & 1 & 1 & -3 & 1 & 1 & -1/3 & -3 \\
 $\mathcal{F}_4$ & -3 & -3 & 1 & 1 & 9 & 9 & -3 & 1 & -3 \\

\hline
\end{tabular}
\caption{The CG coefficient, the parameters $\alpha$, $\beta$ and
$h$ and their ratios for the sixteen categories of neutrino mass
matrices. We have given $\alpha$, $\beta$ and $h$ in the units of
$\left(\frac{2 a_u b_u}{r}\right)$, $\left(\frac{a_u^2}{r}\right)$
and $\left(\frac{a_u c_u}{r}\right)$ respectively.}
\end{center}
\end{table}

\end{document}